\documentclass [12pt]  {article}
\def\vP{\vec{P}\, {}^2}
\begin{document}
\title{%
Doubly Special Relativity with light-cone deformation}
\author{A.\ B\l{}aut\thanks{e-mail address ablaut@ift.uni.wroc.pl},
M.\ Daszkiewicz\thanks{e-mail address marcin@ift.uni.wroc.pl},\,
and J.\ Kowalski--Glikman\thanks{e-mail address
jurekk@ift.uni.wroc.pl}~~\thanks{Partially supported by the   KBN
grant 5PO3B05620}\\  \\ {\em Institute for Theoretical
Physics}\\ {\em University of Wroc\l{}aw}\\ {\em Pl.\ Maxa Borna 9}\\
{\em Pl--50-204 Wroc\l{}aw, Poland}} \maketitle
\begin{abstract}
We propose a new Doubly Special Relativity theory based on the
generalization of the $\kappa$-deformation of the Poincar\'e
algebra acting along one of the null directions. We recall the
quantum Hopf structure of such deformed Poincar\'e algebra and
use it to derive the phase space commutation relations. As in the
DSR based on the standard quantum $\kappa$-Poincar\'e algebra we
find that the space time is non-commutative. We investigate the
fate of the properties of Special Relativity in the null basis:
the split of the algebra of Lorentz and momentum generators into
kinematical and dynamical parts, the action of the kinematical
boost $M^{+-}$, and the emergence of the two dimensional Galilean
symmetry.
\end{abstract}
\clearpage

\section{Introduction}
 The fate of Lorentz symmetry at high energies close to the
Planck scale is recently attracting growing interest. The reason
for this is at least twofold. First, since the classical Lorentz
symmetry group is non-compact we have only limited access to
experimental verification of this symmetry. Second, the observed
anomalies in the ultra-high energy cosmic rays and TeV photons
\cite{astro} might indicate that relativistic kinematics may
deviate from the standard Special Relativistic one when relevant
energies becomes close to the Planck scale.

There are two major approaches to the ``Planck scale
phenomenology'' (see \cite{Amelino-Camelia:2002ar} and
\cite{Amelino-Camelia:2002fw} for reviews and extesive list of
references). In the first one assumes that Lorentz symmetry
becomes broken close to the Planck scale, so that effectively we
have to do with some preferred frame, usually associated with the
cosmic frame. In another approach, dubbed Doubly Special
Relativity \cite{Amelino-Camelia:2000mn},
\cite{Amelino-Camelia:2000ge}, one assumes that Lorentz (and even
Poincar\'e) symmetry is still present, but becomes deformed in the
high energy regime. This approach borrows a lot from previous
investigations on quantum deformation of Poincar\'e Hopf algebra,
undertaken in 1990s (see e.g., \cite{lunoruto}, \cite{maru}).

Presently two Doubly Special Relativity models have received most
attention\footnote{As shown in the series of papers
\cite{Kowalski-Glikman:2002ft}, \cite{Kowalski-Glikman:2002jr},
 \cite{Kowalski-Glikman:2002we} there are in fact infinitely many DSR
theories.}. In the first, called DSR1  formulated in
\cite{Kowalski-Glikman:2001gp} and \cite{Bruno:2001mw} the
modified dispersion relation takes the form
\begin{equation}\label{1}
 m^2 = \left(2\kappa \sinh \left(\frac{P_0}{2\kappa}\right)\right)^2 - \vec{P}\,{}^2\, e^{P_0/\kappa}.
\end{equation}
while in the DSR2, proposed in \cite{Magueijo:2001cr} it reads
\begin{equation}\label{2}
 m^2 = \frac{P_{0}^2 - \vec{P}{}^2}{\left(1- \frac{P_0}\kappa\right)^2},
\end{equation}
Let us note that the deformation in the dispersion relations
(\ref{1}), (\ref{2}) act in the ``energy'' direction. This is a
general feature of all DSR theories considered so far, and a
natural question arises as to whether it is possible to construct
DSR theories with different kind of deformations. For example is
it possible to construct a DSR theory with deformation along
``light cone'' corresponding to the front form dynamics in the
famous classification of Dirac \cite{Dirac1949} (see also e.g.
\cite{Leutwyler1978}, \cite{Heinzl2000} for extensive up-to-date
discussion and list of references.) The answer to this question
happens to be affirmative \cite{Ballesteros:mi},
\cite{kosinski1995},  \cite{przanowski1996},
\cite{Lukierski:2002ii} and we devote this note to investigate
most fundamental properties of such theories.

Our motivations to investigate light-cone DSR are manyfold. First,
such  DSR theories differ in many respects from the theories
analyzed so far and therefore are of interest on its own. Moreover
the standard light-cone dynamics has a number of distinctive
features which, in the deformed case may shed some light on
generic structure of Doubly Special Relativity. Among them one
should mention the rational relation between energy and momentum
on-shell $P_- = \frac{\vP + m^2}{P_+}$ as compared to $P_0
=\sqrt{\vP + m^2}$ of standard ``instant'' formulation of Special
relativity, special structure of boosts of which one is a
kinematical operator acting along the constant ``light-cone time''
surface, and the emergence of Galilean symmetry  as a subgroup of
the whole Poincar\'e algebra (and not the reduction of the latter,
as in the standard case.) Last but not least, it were intuitions
from the light cone dynamics that lead Susskind \cite{Susskind} to
formulate his version of the holographic principle, so one may
hope that proper understanding of the deformed light-cone dynamics
may lead to better understanding of this principle.

\section{DSR phase space in light-cone coordinates}

Our starting point of the construction of the light-cone DSR
theory will be the light-cone quantum $\kappa$-Poincar\'e algebra.
This kind of algebras have been investigated first in
\cite{Ballesteros:mi}, whose construction has been incorporated
into the generalized $\kappa$-Poincar\'e framework in
\cite{kosinski1995} and \cite{przanowski1996}. As usual in the
light-cone formulation momenta has components $P_+$, $P_-$ along
the light-cone and the transverse ones $P_i$\ \footnote{In what
follows we will use the following index convention: Greek indices
run through $+,-,1,2$, small Latin indices denote transverse
directions $1,2$, while capital ones run through $-,1,2$.}. The
metric tensor in such parameterization has the form
$$
\label{me} g_{\mu\nu}= g^{\mu\nu}=\left(
\begin{array}{cccc}
  0 & 1  & 0  & 0 \\
  1 & 0 & 0  & 0 \\
  0 & 0  & -1 & 0 \\
  0 & 0  & 0  & -1
\end{array}
\right)
$$
The action of Lorentz generators $M^{\mu\nu}$ satisfies the
standard {\em undeformed} algebra
\begin{equation}\label{3}
\lbrack  M^{\mu \nu } , M^{\lambda \sigma } \rbrack   =  i(g^{\mu
\sigma } M^{\nu \lambda }-g^{\nu \sigma }M^{\mu \lambda }+g^{\nu
\lambda } M^{\mu \sigma }-g^{\mu \lambda }M^{\nu \sigma }),
\end{equation}
and the momenta commute
\begin{equation}\label{4}
\lbrack P_\mu, P_\nu \rbrack   =0.
\end{equation}
As usual in DSR the action of Lorentz generators on momenta is
deformed with the scale of deformation given by the parameter
$\kappa$ of dimension of mass.
\begin{eqnarray}
\lbrack  M^{ij},P_+ \rbrack  &=&  \lbrack  M^{i-},P_+ \rbrack =0\nonumber \\
\lbrack  M^{i+},P_+ \rbrack  &=& i P^i\nonumber \\
\lbrack  M^{-+},P_+ \rbrack  &=& i \kappa \left(1- e^{-P_+/\kappa}\right)\nonumber \\
\lbrack  M^{ij},P_- \rbrack  &=& 0\nonumber \\
\lbrack  M^{i+},P_- \rbrack  &=& -\frac{i}\kappa P^iP_-\nonumber \\
\lbrack  M^{i-},P_- \rbrack  &=& i P^i\nonumber \\
\lbrack  M^{-+},P_- \rbrack  &=& -i P_- -\frac{i}{2\kappa} \vP\nonumber \\
\lbrack  M^{ij},P_k \rbrack  &=& i\left(\delta ^j_{\ k}P^i-\delta ^i_{\ k}P^j\right)\nonumber \\
\lbrack  M^{i+},P_k \rbrack  &=& -i\delta ^i_{\ k} \left(P_-\,
e^{-P_+/\kappa} + \frac{1}{2\kappa}\vP\right)
- \frac{i}{\kappa}P^iP_k\nonumber \\
\lbrack  M^{i-},P_k \rbrack  &=& -i \kappa \delta ^i_{\ k} \left(1- e^{-P_+/\kappa}\right)\nonumber \\
\lbrack  M^{-+},P_k \rbrack  &=& - i P_k \left(1-
e^{-P_+/\kappa}\right),\label{5}
\end{eqnarray}
where we define $\vP = P_1^2 + P_2^2$. The Casimir  (dispersion
relation) for this algebra reads
\begin{equation}\label{6}
 m^2=4\kappa
P_{-}e^{P_{+}/2\kappa}\sinh{\frac{P_{+}}{2\kappa}}-\left(P_1^2+P_2^2\right)e^{P_{+}/\kappa}
\end{equation}
which, of course, reduces (as well as the algebra above) to the
standard relation of the null-frame Special Relativity in the
$\kappa\rightarrow\infty$ limit. Note that in the large $P_+$
limit the dispersion relation becomes in the leading order
\begin{equation}\label{a}
\vP = 2\kappa P_-
\end{equation}
independently of mass, as compared with the relation
$$
\vP + m^2 = 2P_+ P_-
$$
of the undeformed case.

The fact that the algebra (\ref{5}) can be extended to a quantum
Hopf algebra makes it possible to use the co-product of the
latter to construct the commutators algebra of the whole  the
phase space of the system. This procedure consists first of
defining pairing between Lorentz generators and momenta position
$X^\mu$. To do that one takes the Hopf algebra co-products
\begin{equation}\label{7}
\begin{array}{rcl}
\Delta(P_+)&=&1\otimes P_+ +P_+\otimes 1\\
\Delta(P_I)&=&P_I\otimes e^{-P_+/\kappa}+1\otimes P_I\\
\Delta(M^{IJ})&=&1\otimes M^{IJ}+M^{IJ}\otimes 1\\
\Delta(M^{I+})&=&M^{I+}\otimes e^{-P_+/\kappa}+1\otimes M^{I+}-\frac{1}{\kappa}M^{IJ}\otimes P_J\\
\Delta(X^{\mu})&=&\Lambda^{\mu}{}_{\nu}\otimes X^{\nu}+X^{\mu}\otimes 1\\
\Delta(\Lambda^{\mu}{}_{\nu})&=&\Lambda^{\mu}{}_{\rho}\otimes
\Lambda^{\rho}{}_{\nu}\\
\end{array}
\end{equation}
and then defines the pairing as follows
\begin{equation}\label{8}
\begin{array}{rcl}
<P_{\mu},X^{\nu}>&=&i\delta_{\mu}^{\nu}\\
<\Lambda^{\mu}{}_{\nu},M^{\alpha\beta}>&=&i
\left(g^{\alpha\mu}\delta^{\beta}_{\nu}-g^{\beta\mu}\delta^{\alpha}_{\nu}\right)\\
<\Lambda^{\mu}{}_{\nu},1>&=&\delta^{\mu}_{\nu}
\end{array}
\end{equation}
In the final step one uses eqs.~(\ref{8}) to derive relevant
commutators according to the general formula
\begin{equation}\label{9}
 \left[{\cal Q}, {\cal R}\right] = {\cal R}_{(1)}\, <{\cal Q}_{(1)},{\cal R}_{(2)}>\, {\cal Q}_{(2)}
-{\cal R}{\cal Q},
\end{equation}
where we used a natural notation for co-product $\Delta\, {\cal R}
= \sum {\cal R}_{(1)}\otimes {\cal R}_{(2)}$. The resulting
non-vanishing phase space commutators read
\begin{equation}\label{10}
\left[X^{+},X^{i}\right]=\frac{i}{\kappa}X^i\quad
\left[X^{+},X^{-}\right]=\frac{i}{\kappa}X^-
\end{equation}
\begin{equation}\label{11a}
\left[P_{i},X^{j}\right]=-i\delta_{i}^{j}\quad
\left[P_{i},X^{+}\right]=\frac{i}{\kappa}P_i
\end{equation}
\begin{equation}\label{11b}
\left[P_{-},X^{-}\right]= -i\quad
\left[P_{-},X^{+}\right]=\frac{i}{\kappa}P_-\quad
\left[P_{+},X^{+}\right]=-i
\end{equation}
The transformation of positions under action of Lorentz generators
are given by
\begin{equation}\label{12a}
\left[X^{\mu},M^{K+}\right]=i\left(g^{K\mu}X^{+}-g^{+\mu}X^{K}\right)
-\frac{i}{\kappa}\left(\delta_{+}^{\mu}M^{K+}+\delta_{L}^{\mu}M^{KL}\right)
\end{equation}
\begin{equation}\label{12b}
\left[X^{\mu},M^{KL}\right]=i\left(g^{K\mu}X^{L}-g^{L\mu}X^{K}\right)
\end{equation}

To conclude this technical section stressing that the phase space
algebra constructed above can be altered by redefinition of
momenta of the form
\begin{equation}\label{13}
\begin{array}{rcl}
P_+ &\rightarrow& {\cal P}_+\, \left(P_+, P_-, \vP\right) \nonumber \\
P_- &\rightarrow& {\cal P}_-\, \left(P_+, P_-, \vP\right) \nonumber \\
P_- &\rightarrow& {\cal P}_i\, \left(P_+, P_-, \vP, P_i\right) =
A\left(P_+, P_-, \vP\right)\, P_i
\end{array}
\end{equation}
This redefinition is restricted only by the natural requirement
that in the $\kappa\rightarrow\infty$ limit $P_\mu ={\cal P}_\mu$,
which means that in this limit all the light-cone DSR theories
reduce to Special Relativity in the null frame.

\section{Properties of the phase space algebra}

Let us now discuss the properties of the deformed phase space
derived in the previous section. We will focus our attention on
the fate of three distinguishing properties of Special Relativity
formulated in null frame: split of the Poincar\'e generators into
seven kinematical and three dynamical ones, the simple action of
the boost in the third direction $M^{-+}$ and the emergence of the
Galilei algebra as a subalgebra of the Poincer\'e light-cone
algebra. Before doing that, let us briefly describe these three
features.

In his seminal paper \cite{Dirac1949} Dirac considered the
possible forms of initial surfaces for relativistic dynamics.
Given an initial surface it is important to know which Poincar\'e
symmetry generators act along the surface (these are called
kinematical) and which in the orthogonal direction (since these
generators change the value of the coordinate orthogonal to the
surface -- the time variable -- they are called dynamical.) In the
case of constant time initial surface (instant form Special
Relativity), it is well known that there are six kinematical
(spacial momenta and rotations) generators and four dynamical ones
(boosts and energy generator). In the null frame  Special
Relativity, when the initial surface is null it turns out that
there are seven kinematical generators (it can be shown that this
is the maximal possible number of them) and three kinematical
ones.

The generator that is dynamical in the instant form, and becomes
kinematical in the null form is the boost along the third axis
$M^{+-}$. Moreover this generator acts as a mere rescaling of
positions and momenta, to wit
$$
X^+ \rightarrow e^\xi\,X^+, \quad X^- \rightarrow e^{-\xi}\,X^-,
\quad P^+ \rightarrow e^\xi\,P^+, \quad P^- \rightarrow
e^{-\xi}\,P^-,
$$
where $\xi$ is the boost parameter. Clearly these transformation
leaves the surface $X^+=0$ (or $X^-=0$) invariant, so indeed
$M^{+-}$ is kinematical.

The last feature of the null frame form is that the subalgebra of
the whole Poincar\'e algebra of the longitudinal symmetry
generators is isomorphic to the Galilei \cite{SBH} algebra with
the following identification: $P^i$ correspond to Galilei momenta,
$M^{ij}$ to rotations, $P^-$ to Hamiltonian, while $P^+$ is a
central charge of the subalgebra, corresponding to the Galilei
mass (see below.) This situation should be compared with the
instant form Special Relativity, where the Galilei algebra arises
only as an appropriate limit of the Poincar\'e algebra.

To conclude these introductory remarks, let us note that contrary
to the Special Relativistic case there is no $+ \leftrightarrow -$
symmetry between coordinates $X^+$ and $X^-$ (and momenta $P^+$
and $P^-$), and we must therefore consider the cases of null
surfaces $X^+=0$ and $X^-=0$ separately. As it will turn out these
surfaces have very different properties.

\subsubsection*{Kinematical generators}

In the first case, in order to find out which generators are
kinematical and which dynamical we must consider generators of the
stability group of the surface $X^+=0$. Consider first the algebra
(\ref{12a}), (\ref{12b}). We have
\begin{equation}\label{13a}
\left[X^{+},M^{K+}\right]= -\frac{i}{\kappa}M^{K+}
\end{equation}
\begin{equation}\label{13b}
\left[X^{+},M^{KL}\right]=i\left(g^{K+}X^{L}-g^{L+}X^{K}\right)
\end{equation}
From  (\ref{13b}) we learn that  $M^{ij}$ is kinematical, as in
the standard case. However $M^{+i}$ and $M^{+-}$ have
non-vanishing commutator (\ref{13a}) and therefore are not
kinematical. The inspection of the algebra (\ref{11a}),
(\ref{11b}) shows that also the momenta $P_i$, $P_-$ are not
kinematical.

Consider therefore  the null surface $X^-=0$. The algebra
(\ref{12a}), (\ref{12b}) reads in this case
\begin{equation}\label{14a}
\left[X^{-},M^{K+}\right]=-i X^{K} -\frac{i}{\kappa}M^{K-}
\end{equation}
\begin{equation}\label{14b}
\left[X^{-},M^{KL}\right]=0
\end{equation}
We see that now the kinematical generators are $M^{ij}$, $M^{-+}$,
$M^{i-}$. From (\ref{11a}), (\ref{11b}) we see that also the
momenta $P_i$, $P_-$ are kinematical. Therefore in this case we
have  as many kinematical  generators as in Special Relativity.

To conclude, for the initial surface $X^+=0$ the number of
kinematical generators is small, but the coordinates on the
surface are commutative. It is light-cone time $X^+$ that is
non-commutative, and the light-cone energy can be associated with
the $P_+$ generator. For the initial surface $X^-=0$ the number of
kinematical generators is exactly equal to the corresponding
number in Special Relativity, but, the coordinates on the surface
are not commutative. In this case $X^-$ and $P_-$ play a role of
time and energy, respectively.

\subsubsection*{The $M^{-+}$ boost}

 To see what is the behavior of
the $M^{-+}$ boost for deformed algebra, let us first consider the
commutators with momenta (\ref{5}) and infer from them
\begin{eqnarray}
\frac{ d\, P_+}{d\xi}   &=&  \kappa \left(1- e^{-P_+/\kappa}\right)\nonumber \\
\frac{ d\, P_-}{d\xi}  &=& - P_- -\frac{1}{2\kappa} \vP\nonumber \\
\frac{ d\, P_k}{d\xi}  &=& - P_k \left(1-
e^{-P_+/\kappa}\right)\nonumber
\end{eqnarray}
where $\xi$ is the rapidity parameter. The solution of these
equations with the initial condition $P_\mu(0) = \pi_\mu$ reads
$$
 P_+(\xi) = \kappa\, \log\left[1 +  C\, e^\xi\right],
 \quad P_i(\xi) = \frac{(C+1)\, \pi_i}{1 + C\, e^\xi},
 $$
 \begin{equation}\label{15}
  P_-(\xi) = e^{-\xi} \left[  \pi_- - \frac{(C+1)\, \vec{\pi}\,{}^2}{2\kappa
  C}+\frac{(C+1)^2\, \vec{\pi}\,{}^2}{2\kappa C(1 + C\,
  e^\xi)}\right]
  \end{equation}
where $C = e^{\pi_+/\kappa} -1$. This last equation can be
simplified if one uses the dispersion relation satisfied by the
initial values of momenta
$$
 4\kappa
\pi_{-}e^{\pi_{+}/2\kappa}\sinh{\frac{\pi_{+}}{2\kappa}}-\vec{\pi}\,{}^2\,
e^{\pi_{+}/\kappa}=m^2.
$$
One gets
\begin{equation}\label{15a}
P_-(\xi) = e^{-\xi} \left[ \frac{m^2}{2\kappa
  C}+\frac{(C+1)^2\, \vec{\pi}\,{}^2}{2\kappa C(1 + C\,
  e^\xi)}\right]
\end{equation}

Expanding to the lowest order in $\kappa$ we get the standard
relations
\begin{equation}\label{16}
P_+(\xi) \sim\pi_+ \, e^\xi, \quad P_-(\xi) \sim \pi_-\, e^{-\xi},
\quad P_i(\xi) \sim 0
\end{equation}
as it should be. The expansion for large boosts
$\xi\rightarrow\infty$ gives instead (in the massless case, for
simplicity)
\begin{equation}\label{17}
P_+(\xi) \sim  \xi ,
  \quad P_i(\xi)\sim
e^{-\xi},\quad P_-(\xi) \sim  e^{-2\xi}.
\end{equation}
This expansion is consistent with relation eq.~(\ref{a}). Note
that while the behavior of $P_-(\xi)$ is similar to the standard
one, $P_+(\xi)$ grows with $\xi$ much slower than in Special
Relativity.

\subsubsection*{Galilean symmetry}

As mentioned above, the last interesting feature of the Special
Relativity in the null frame is that there exist two subalgebras
of the full Poincar\'e algebra that are isomorphic to the algebra
of the two-dimensional Galilei group. These subalgebras consist of
the rotation $M^{ij}$, Galilean boosts $G_\pm^i = M^{\pm i}$, and
$P_\mu$ and  read
\begin{eqnarray}
\lbrack  M^{ij},G_\pm^k\rbrack  &=& i\left(g^{ik}G_\pm^j - g^{jk}G_\pm^i\right)\nonumber \\
\lbrack  M^{ij},P_k\rbrack  &=& i\left(\delta^j_{\ k}P^i-\delta ^i_{\ k}P^j\right)\nonumber \\
\lbrack  G_\pm^i,P_+ \rbrack  &=& -i\delta^\pm_+\, P^i\nonumber \\
\lbrack  G_\pm^i,P_-\rbrack  &=& -i\delta^\pm_-\, P^i\nonumber \\
\lbrack  G_\pm^i,P_k\rbrack  &=& -i \delta^i_{\ k}\,
P_{\mp}\label{19}
\end{eqnarray}
Note that in the `--' case $P_+$ is a central charge of the
algebra and can be therefore identified with (twice) the Galilean
mass, while $P_-$ is  the Hamiltonian (in the `+' case the mass is
identified with $P_-$, and Hamiltonian with $P_+$.)

Inspection of the algebra (\ref{5}) shows that in the deformed
algebra this doubly structure is lost. Indeed, in this case only
$P_+$ can be central charge, as it commutes with $G^i_-$. In this
way we obtain the deformed Galilei algebra, being a subalgebra of
the deformed Poincar\'e algebra (\ref{5}), to wit
\begin{eqnarray}
\lbrack  M^{ij},G^k\rbrack  &=& i\left(g^{ik}G_\pm^j - g^{jk}G_\pm^i\right) \\
\lbrack  M^{ij},P_k\rbrack  &=& i\left(\delta^j_{\ k}P^i-\delta ^i_{\ k}P^j\right)\nonumber \\
\lbrack  G^i,P_-\rbrack  &=& -i\delta^\pm_-\, P^i\nonumber \\
\lbrack  G^i,P_k\rbrack  &=& i \kappa \delta ^i_{\ k} \left(1-
e^{-P_+/\kappa}\right)\label{18}
\end{eqnarray}
where we used the notation $G^i=G_-^i$. Since $P_-= P^+$ is the
Hamiltonian, it corresponds to the initial null surface defined by
$X^-=0$, because in this case $X^-$ takes a role of time.

\section{Conclusions}

In this paper we investigated basic properties of a new Doubly
Special Relativity in which the deformation acted along the light
cone. This theory differs in many aspects from the DSR theories
considered so far, and in our opinion deserves farther studies.

However it is our view that the final judgment of which theory is
correct is provided by its experimental verification. In the case
of the DSR theories one of the few experiments that can be
performed in a near future are ``time of flight experiments'',
i.e., measurements of dependence of the speed of light on momentum
carried by photons (see, e.g., \cite{Amelino-Camelia:1999zc} and
references therein.) Therefore, it is important to investigate in
details the issue of velocity in DSR theories. We will address
this problem in a future publication.

\section*{Acknowledgement}

We would like to thank J. Lukierski for discussion and bringing
the paper \cite{Lukierski:2002ii} to our attention.

\end{document}